\documentclass[conference]{IEEEtran}
\IEEEoverridecommandlockouts
\usepackage{cite}
\usepackage{amsmath,amssymb,amsfonts}
\usepackage{algorithmic}
\usepackage{graphicx}
\usepackage{textcomp}
\usepackage{xcolor}
\usepackage{booktabs}
\def\BibTeX{{\rm B\kern-.05em{\sc i\kern-.025em b}\kern-.08em
    T\kern-.1667em\lower.7ex\hbox{E}\kern-.125emX}}

\newcommand{\eg}{e.\,g.,\ }
\newcommand{\ie}{i.\,e.,\ }

\usepackage{soul}
\usepackage{xcolor}
\definecolor{newgreen}{RGB}{34,139,34}

\newcommand{\sstitle}[1]{\smallskip\noindent\textbf{#1.\/}}

\begin{document}

\title{Diff-ETS: Learning a Diffusion Probabilistic Model for Electromyography-to-Speech Conversion
\thanks{Corresponding author: Zhao Ren (zren@uni-bremen.de).\\
This research is funded by the Deutsche Forschungsgemeinschaft (DFG, German Research Foundation) through the project ``MyVoice: Myoelectric Vocal Interaction and Communication Engine'' (460884988).
This work has been submitted to the IEEE for possible publication. Copyright may be transferred without notice, after which this version may no longer be accessible.}
}

\author{
\centering
    \IEEEauthorblockN{
        Zhao Ren\IEEEauthorrefmark{1},
        Kevin Scheck\IEEEauthorrefmark{1}, 
        Qinhan Hou\IEEEauthorrefmark{2}, 
        Stefano van Gogh\IEEEauthorrefmark{2}, 
        Michael Wand\IEEEauthorrefmark{2}\IEEEauthorrefmark{3},    
        Tanja Schultz\IEEEauthorrefmark{1}    
    }
    \IEEEauthorblockA{\IEEEauthorrefmark{1} Cognitive Systems Lab, University of Bremen, Bremen, Germany}
    \IEEEauthorblockA{\IEEEauthorrefmark{2} Istituto Dalle Molle di studi sull'intelligenza artificiale (IDSIA USI-SUPSI), Lugano, Switzerland}
    \IEEEauthorblockA{\IEEEauthorrefmark{3} Institute for Digital Technologies for Personalized Healthcare, SUPSI, Lugano, Switzerland}
}


\maketitle

\begin{abstract}
Electromyography-to-Speech (ETS) conversion has demonstrated its potential for silent speech interfaces by generating audible speech from Electromyography (EMG) signals during silent articulations. ETS models usually consist of an EMG encoder which converts EMG signals to acoustic speech features, and a vocoder which then synthesises the speech signals. Due to an inadequate amount of available data and noisy signals, the synthesised speech often exhibits a low level of naturalness. In this work, we propose Diff-ETS, an ETS model which uses a score-based diffusion probabilistic model to enhance the naturalness of synthesised speech. The diffusion model is applied to improve the quality of the acoustic features predicted by an EMG encoder. In our experiments, we evaluated fine-tuning the diffusion model on predictions of a pre-trained EMG encoder, and training both models in an end-to-end fashion. We compared Diff-ETS with a baseline ETS model without diffusion using objective metrics and a listening test. The results indicated the proposed Diff-ETS significantly improved speech naturalness over the baseline.
 


\end{abstract}


\section{Introduction}
Speech, a key mode of human communication, is a complex procedure produced by the brain and muscles, including the larynx, tongue, lips, and jaw~\cite{schultz2017biosignal,kim2023diff}. Communication can be extremely challenging when individuals cannot produce audible speech, negatively affecting their life quality.
In this regard, silent speech interfaces (SSIs) can assist humans to silently communicate by synthesising audible speech from biosignals~\cite{denby2010silent,schultz2017biosignal}.
SSIs can not only help humans with speech impairments (\eg laryngectomy patients), but also support private conversations in public environments and effective interactions in noisy conditions~\cite{schultz2017biosignal,maier-hein2005speech,ren2023anoverview}. More recently, Electromyography (EMG)-to-Speech (ETS) conversion shows its potential in SSIs by synthesising audible speech from facial surface electromyographic signals~\cite{diener2015direct}. Particularly, ETS conversion holds promise for synthesising speech during silent articulation, as facial EMG signals contain the movement information of human articulatory muscles~\cite{denby2010silent,wand2011session}. 

Different from EMG-based speech recognition, which predicts sentences from EMG in early studies~\cite{wand2011session,schultz2010modeling}, ETS conversion has been demonstrated to be promising to produce acoustic speech directly from EMG~\cite{scheck2023stream,ren2023self,scheck2023multi}. An ETS model usually consists of an EMG encoder to predict acoustic features from EMG signals, and a vocoder to synthesise speech from these predictions~\cite{scheck2023multi,gaddy2021improved}. Direct speech synthesis from EMG can preserve more communication-related information (\eg paralinguistics) than EMG-based speech recognition. 
Nonetheless, due to the low amount of available data or no ground-truth (GT) audio during silent speech, the ETS synthesis often lacks in speech naturalness.
This could potentially limit the user acceptance for EMG-driven SSIs.

Diffusion probabilistic (\ie diffusion-based) models have recently improved the speech naturalness in various synthesis tasks, \eg text-to-speech (TTS)~\cite{jeong2021diff,popov2021grad} and voice conversion~\cite{choi23d_interspeech}. Specifically, a forward process simulates a random walk in the latent space by progressively adding Gaussian noise to the original data, and a reverse process aims to denoise and reconstruct the original data in multiple time steps~\cite{ho2020denoising,kim2023diff}. Diffusion-based models can effectively generate natural speech through their forward and reverse steps~\cite{jeong2021diff,popov2021grad,huang2022prodiff}. 
We propose a diffusion-based model for ETS, namely \textit{Diff-ETS}, including an encoder, a diffusion probabilistic model, and a vocoder. Diff-ETS is found to successfully improve the speech naturalness of ETS without diffusion. Furthermore, the Diff-ETS is flexible to set the reverse time steps at the inference stage, enabling the trade-off between latency and performance.
To the best of the authors' knowledge, this work proposes the first usage of diffusion-based models for ETS.

\sstitle{Related Work}
Various diffusion-based models have been applied to improve speech naturalness in the task of TTS, which is relevant to our ETS work. 
The study in~\cite{popov2021grad} proposed a score-based diffusion probabilistic model that estimates gradients of log-density of noisy data for TTS. This diffusion model was further improved by a phoneme classifier, aiming to reduce the pronunciation error when using untranscribed speech~\cite{kim2022guided}. Another study~\cite{jeong2021diff} applied a denoising diffusion probabilistic model with likelihood-based optimisation for TTS. The score-based diffusion model in~\cite{popov2021grad} showed better naturalness than the denoising diffusion probabilistic model in~\cite{jeong2021diff}.   
To improve the speech naturalness of ETS, the study in~\cite{scheck2023stream} trained a modified HiFi-GAN vocoder together with an EMG encoder in an E2E manner. Different from training the vocoder, we improve the naturalness via Diff-ETS, a score-based diffusion probabilistic model, which has also the potential to be combined with the training of the vocoder.

\section{Methodology}
\begin{figure}
    \centering
    \includegraphics[width=.49\textwidth, trim={3cm 5cm 5cm 2cm}, clip]{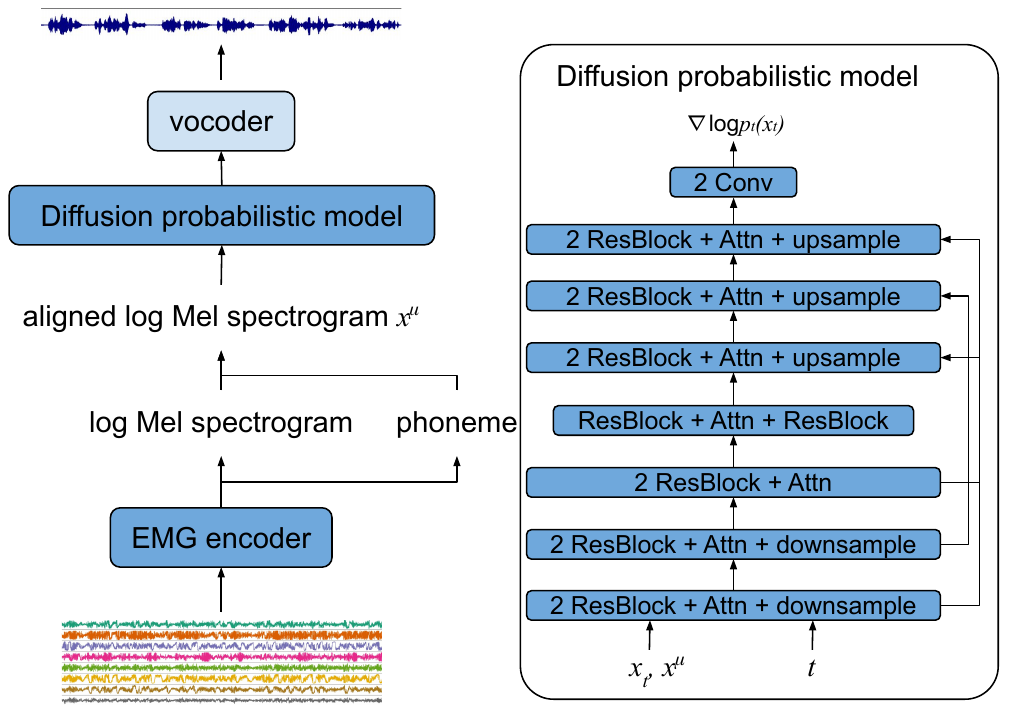}
    \caption{The proposed Diff-ETS framework for ETS. The deep blue blocks are trainable and the light blue block of the vocoder is frozen. ResBlock: Residual blocks, Attn: Attention, Conv: Convolutional layers.}
    \label{fig:diffusion}
\end{figure}

The proposed framework in this work includes an EMG encoder, a diffusion probabilistic model, and a vocoder (see Fig.~\ref{fig:diffusion}). Given an EMG signal, an EMG encoder is trained to predict a log Mel spectrogram and frame-wise phoneme targets. The log Mel spectrogram is then aligned to have the same length as that of the log Mel spectrogram extracted from corresponding audible speech using Dynamic Time Warping (DTW). Next, a score-based diffusion probabilistic model is trained to enhance the log Mel spectrogram. Lastly, a pre-trained vocoder synthesizes the speech signal from the enhanced Mel spectrogram. Notably, the alignment of the log Mel spectrograms is only applied during the training procedure, as the training of the diffusion model requires the inputs and the log Mel spectrograms of the GT (\ie corresponding audible speech) to have the same shape. The log Mel spectrograms predicted by the EMG encoder are fed into the diffusion model without alignment at the inference stage. 

\subsection{EMG Encoder and Speech Vocoder}
In this work, we apply the structure of the EMG encoder proposed in~\cite{gaddy2021improved}. The encoder consists of a set of residual convolutional neural networks (CNNs) and a transformer model, predicting log Mel spectrograms and frame-wise phonemes. Additionally, the speech vocoder is a pre-trained HiFi-GAN model~\cite{kong2020hifi}, which is herein frozen during training.

\subsection{Score-based Diffusion Probabilistic Model}
We use a score-based diffusion probabilistic model between the encoder and the vocoder to increase the speech naturalness. We denote a log Mel spectrogram predicted by the EMG encoder as $x^{\mu}$, and the log Mel spectrogram of the GT is defined by $x_0$.  Herein, the aligned log Mel spectrogram $x^{\mu}$ is used as the input of the diffusion model during training, and the unaligned Mel spectrogram is employed for inference. In the following, $x^{\mu}$ is used to denote the diffusion model input for simplification.

\subsubsection{Forward process}
\label{sec:diffusion_forward}
Given $x^{\mu}$, the forward diffusion process can transform it to Gaussian noise in a sequential stochastic process. At a time step $t$, $t\in[1;T]$, the deviation of the corresponding data $x_t$~\cite{popov2021grad} is defined by 
\begin{equation}
\label{eq:forward1}
    dx_t=\frac{1}{2}(x^\mu-x_t)s_t dt + \sqrt{s_t}dW_t,
\end{equation}
where $s_t$ is the noise schedule and $W_t$ denotes the standard Brownian motion. The study in~\cite{popov2021grad} proved that $(x_t|x_0) \xrightarrow{d}\mathcal{N}(x^\mu, I)$ where $I$ is an identity matrix, when $t$ is infinite. In this regard, $x_t$ obeys the Gaussian noise of $(x^\mu, I)$ and can be obtained without $x_0$.

\subsubsection{Reverse process}
The reverse diffusion process aims to recover the clean data from Gaussian noise by approximating the forward process. Herein, we use the ordinary differential equation, which has been effective for TTS in~\cite{popov2021grad}:
\begin{equation}
\label{eq:reverse1}
    dx_t=\frac{1}{2}(x^\mu-x_t-\nabla\log p_t(x_t))s_tdt.
\end{equation}
Given $dx_t$ computed by~(\ref{eq:reverse1}), $x_t$ can be reversed to approach $x_0$ step-by-step. The $\nabla\log p_t(x_t)$, the gradient of the log density of the noisy data, can be predicted by a trainable neural network $f(x_t, x^\mu,t)$. The architecture of $f(x_t, x^\mu,t)$ is a U-Net with several residual blocks, attention layers, downsampling and upsampling operations (see the right part of Fig.~\ref{fig:diffusion}). The detailed model structure is described in Section~\ref{sec:exp}.

\subsection{Model Training}
In this work, there is flexibility to either i) train the encoder and then train the diffusion model with a frozen encoder, or ii) train the encoder and the diffusion model in an end-to-end (E2E) manner.
In the E2E training, the loss function is computed by 
\begin{equation}
\label{eq:e2e}
\mathcal{L} = \mathcal{L}_{enc} + \lambda_d\mathcal{L}_d, 
\end{equation}
where $\mathcal{L}_{enc}$ is the loss function of the encoder, $\mathcal{L}_d$ means the loss function of the diffusion model, and $\lambda_d$ denotes a constant. 

\subsubsection{EMG encoder}
To train an encoder, $\mathcal{L}_{enc}$ is calculated by $\mathcal{L}_{mel} +\lambda\mathcal{L}_{phone}$, where $\mathcal{L}_{mel}$ is for predicting log Mel spectrograms, $\mathcal{L}_{phone}$ aims to predict phonemes, and $\lambda$ is a constant. 
EMG and audio signals of audible speech are synchronous when they are recorded together.
Therefore, we calculate $\mathcal{L}_{mel}$ by a pairwise 2-norm distance, and $\mathcal{L}_{phone}$ is a cross-entropy loss function using pairwise phoneme predictions from EMG and phoneme labels of audio. 
For EMG signals recorded with silent speech, the lengths of predicted log Mel spectrograms may differ from those of the GT log Mel spectrograms. Thus, the predicted data is aligned based on a cost matrix calculated by $\mathcal{L'}_{mel} +\lambda\mathcal{L'}_{phone}$ \cite{gaddy2021improved}. $\mathcal{L'}_{mel}$ represents the distance between each two frames of the EMG predictions and the GT Mel spectrograms. $\mathcal{L'}_{phone}$ are the phoneme log probabilities of the EMG encoder for the GT phoneme labels. DTW is then applied to the cost matrix, producing the alignment between the silent EMG and the speech targets. For silent speech, the cost matrix values on the DTW path are summed as $\mathcal{L}_{enc}$.

\subsubsection{Diffusion probabilistic model}
The diffusion model trains $f(x_t, x^\mu,t)$ to predict $\nabla\log p_t(x_t)$. Based on (\ref{eq:forward1}), $x_t$ can be obtained by
\begin{equation}
\label{eq:xt_train}
    x_t = (I-e^{-\frac{1}{2}\int_0^t s_mdm})x^\mu + e^{-\frac{1}{2}\int_0^t s_mdm}x_0 + g_t,
\end{equation}
where $g_t$ is sampled from $\mathcal{N}(0, \eta_t I)$, $\eta=1-e^{-\int_0^t s_mdm}$. The $\nabla\log p_t(x_t)$ conditioned on $x_0$ is then represented by $\nabla\log p_{0t}(x_t|x_0) = - g_t (\eta_t I)^{-1}$.
The loss function of the diffusion model is calculated by 
\begin{equation}
\begin{aligned}
    \mathcal{L}_d &= \mathrm{E}_{x_0,t}(\mathrm{E}_{g_t}(||f(x_t, x^\mu,t)-\nabla\log p_{0t}(x_t|x_0)||^2_2)) \\
    &= \mathrm{E}_{x_0,t}(\mathrm{E}_{g_t}(||f(x_t, x^\mu,t)+g_t(\eta_t I)^{-1})||^2_2)).
\end{aligned}
\end{equation}
Notably, $x_t$ is obtained by~(\ref{eq:xt_train}) for model training only. At the inference stage, $x_t$ can be calculated without $x_0$, as aforementioned in Section~\ref{sec:diffusion_forward}. We sample $x_t$ from $\mathcal{N}(x^\mu, \theta ^{-1}I)$ as the temperature $\theta$ was shown to be helpful to improve the speech quality in~\cite{popov2021grad}.

\section{Experiments}
\subsection{Database}
To validate the proposed approach, we use the open-vocabulary EMG database collected in~\cite{gaddydigital}. It includes 8-channel monopolar EMG signals captured at a sampling rate of 1\,kHz and audio signals at a sampling rate of 16\,kHz, from a single speaker. The database consists of i) $1,588$ utterances of EMG signals and corresponding speech in $7$ sessions when the speaker speaks silently and audibly, and ii) $5,477$ utterances of EMG and audible speech signals in $10$ sessions. 
The data distribution in each session can be found in~\cite{ren2023self}. The training, validation, and test set are the same as in~\cite{gaddydigital}.

\subsection{Experimental Settings}
\label{sec:exp}
\sstitle{Implementation Details} 
For the EMG encoder, the residual CNNs with three residual blocks are followed by a linear layer.
A transformer encoder is further employed after the linear layer. Finally, two linear layers are used to predict log Mel spectrograms and phonemes, respectively. The detailed model hyperparameters can be found in~\cite{gaddy2021improved}.

In the diffusion model, the input channel numbers of the residual blocks are 2, 64, 64, 128, 128, 256, 256, 256, 512, 128, 256, 64, 128, and 2, respectively. The input and output channel numbers of the final two convolutional layers are (2, 2) and (2, 1) respectively. The downsampling operation is a convolutional layer with a stride of 2, and the upsampling is a transposed convolutional layer with a stride of 2.
The Diff-ETS models are initialised by a pre-trained EMG encoder and a pre-trained diffusion model. The encoder is pre-trained on the EMG database for 100 epochs with $\lambda=0.5$ and an Adam optimiser with an initial learning rate of $10^{-4}$. The learning rate is reduced by half when the validation loss does not decrease in 20 epochs. The diffusion model is pre-trained by the Grad-TTS method~\cite{popov2021grad} on the LJSpeech dataset. 

Diff-ETS models are trained in two ways: one involves training the diffusion model with a frozen encoder (\ie diffusion-finetune), and the other is to train the diffusion model and the encoder in an E2E manner (diffusion-E2E). The Diff-ETS models are trained with an Adam optimiser with an initial learning rate of $10^{-4}$.  
The diffusion-finetune model is trained for $2,000$ epochs, as a high number of epochs has led to speech naturalness improvements in Grad-TTS~\cite{popov2021grad}.
The diffusion-E2E model was trained with $\lambda_d=1$ for only 400 epochs, as it overfitted with more epochs in our preliminary study.

\sstitle{Evaluation Metrics} For evaluating the models, we apply both objective and subjective methods. The objective evaluation metrics contain log-spectral distance (LSD)\footnote{https://github.com/haoheliu/ssr\_eval/tree/main}, Frecht audio distance (FAD)\footnote{https://github.com/microsoft/fadtk}, word error rate (WER), and character error rate (CER).
For calculating the WERs and CERs, we transcribe the speech synthesis with the Whisper ``medium.en" model and compare the normalised hypothesis with the normalised GT transcription. 
For the subjective evaluation, we invite $10$ listeners to evaluate the naturalness of 15 randomly selected speech samples from each model and GT audio recordings. The audio samples are presented in random order and the listeners did not know to which model a sample belongs. The speech naturalness is evaluated with scores varying from $1$ to $5$, where $1$ indicates the poorest quality, and $5$ is the best. Based on the ratings, we calculate the mean opinion score (MOS).

\subsection{Results and Discussion}

\begin{figure}
    \centering
    \includegraphics[width=.4\textwidth]{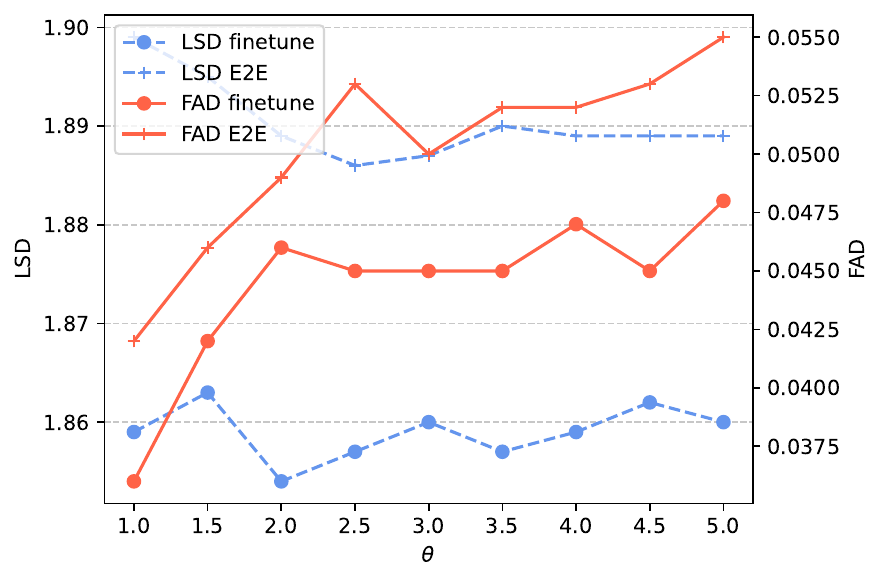}
    \vspace{-15pt}
    \caption{The LSD and FAD of the two Diff-ETS models were evaluated on the test set. The number of reverse time steps $T$ is $50$, and the temperature $\theta$ varies from $1.0$ to $5.0$. LSD: log-spectral distance, FAD: Frecht audio distance. }
    \label{fig:result}
\end{figure}

\begin{table}[]
    \centering
      \footnotesize
 \setlength{\tabcolsep}{4pt} 
    \caption{The performance comparison of Diff-ETS and an encoder, as well as the ground truth (GT).  The number of time steps $T$ in the reverse procedure varies in $\{50, 100, 1,000\}$. A Mann-Whitney U test is used to determine the significance of MOS improvement compared to the encoder (*: $p\leq 0.01$, **: $p\leq 0.001$).}
    \begin{tabular}{lr@{\hskip 0.2in}  rrrrl}
    \toprule
         \textbf{Model} & \textbf{\textit{T}} & \textbf{LSD} & \textbf{FAD} & \textbf{WER} & \textbf{CER} & \textbf{MOS}\\
         \midrule
         GT &-- &-- &-- &2.9 &0.8 & 4.38\\
         Encoder (no diffusion) & -- & 1.869 & 0.089  & \textbf{30.0} & \textbf{15.6} & 3.19\\
         \midrule 
         Diffusion-Finetune &  50 & 1.857& 0.045& 32.4 &17.3 & 3.60*\\
         Diffusion-Finetune & 100 &1.860& 0.042& 32.4 & 17.3 & 3.66**\\
         Diffusion-Finetune & 1,000 &\textbf{1.857}& \textbf{0.042}& 32.1 & 17.0 & \textbf{3.70}**\\
         Diffusion-E2E & 50&1.890  & 0.052& 33.3 & 17.6 &3.59*\\
         Diffusion-E2E & 100 &1.890 & 0.050& 35.2 & 18.6 & 3.57*\\
         Diffusion-E2E & 1,000 &1.888 & 0.048 & 34.1 & 18.7 & 3.58*\\
    \bottomrule
    \end{tabular}
    \label{tab:result}
\end{table}

We compared the LSD and FAD of the diffusion-finetune and diffusion-E2E models using different temperature values $\theta$ in Fig.~\ref{fig:result}. The LSD values were only slightly affected by the temperatures. This indicated that the log Mel spectrograms can be predicted in the reverse process of the diffusion-based models when adding noise with different variances.
The LSD values were slightly higher on the diffusion-finetune model when $\theta$ is smaller than $2.5$. This may arise due to the fact that small $\theta$ values led to larger noise added to the predicted log Mel spectrograms from the EMG encoder.  
The FAD values increased when $\theta$ was larger, which meant that the speech samples predicted by Diff-ETS were still similar to the original GT at a representation-based high level.

We further compared the objective and subjective evaluation metrics of Diff-ETS with different reverse time steps $T$, as well as the GT audio, and the EMG encoder baseline without diffusion in Table~\ref{tab:result}. At the inference stage, the diffusion models herein were applied with $\theta=3.5$. This value was selected using MOS values predicted by a pre-trained UTMOS model\footnote{https://github.com/tarepan/SpeechMOS} in our preliminary study. We observed that the LSD values of the diffusion-finetune models were slightly lower than that of the encoder, while those of the diffusion-E2E models were slightly higher. 
This can be also seen in the WER and CER values, on which the diffusion-E2E models performed slightly worse than others. All Diff-ETS models performed better than the encoder when comparing the FAD values at the representation level. The WER and CER values of all diffusion-finetune and diffusion-E2E models performed slightly worse than the encoder. The reason for this finding might be that only the log Mel spectrograms predicted by the encoder were fed into the diffusion models without considering the predicted phonemes. The diffusion-based models showed significantly better performance on the MOS values for speech naturalness than the encoder in a Mann-Whitney U test. The best performance was a MOS value of $3.70$ ($p \leq 0.001$), obtained by the diffusion-finetune model with $T=1,000$.

\section{Conclusions and Future Work}
This paper proposed a diffusion-based framework called Diff-ETS to enhance the speech naturalness in the task of Electromyography (EMG)-to-Speech (ETS) conversion. The Diff-ETS included two types of training procedures: diffusion-finetune and diffusion-E2E. The diffusion models were added between the EMG encoder and the speech vocoder. The experiments in this work showed a significant improvement in the speech naturalness compared to training an encoder only. The flexibility of the diffusion models allows for the trade-off between latency and performance by setting different reverse time steps.
In future efforts, one can reduce the number of model parameters using various methods, \eg model compression~\cite{merlin2020squeeze} and knowledge distillation~\cite{ren2023fast}, thereby generating speech samples in real-time. Moreover, a diffusion model can be trained together with the encoder and vocoder for further enhancing the speech quality~\cite{scheck2023stream}.


\bibliographystyle{ieeetr}
\bibliography{paper}

\begin{thebibliography}{10}

\bibitem{schultz2017biosignal}
T.~Schultz, M.~Wand, T.~Hueber, K.~D. J., C.~Herff, and J.~S. Brumberg, ``Biosignal-based spoken communication: A survey,'' {\em IEEE/ACM Transactions on Audio, Speech and Language Processing}, vol.~25, no.~12, pp.~2257--2271, 2017.

\bibitem{kim2023diff}
S.~Kim, Y.-E. Lee, S.-H. Lee, and S.-W. Lee, ``{Diff-E: Diffusion-based learning for decoding imagined speech EEG},'' in {\em Proc.\ INTERSPEECH}, (Dublin, Ireland), pp.~1159--1163, 2023.

\bibitem{denby2010silent}
B.~Denby, T.~Schultz, K.~Honda, T.~Hueber, J.~Gilbert, and J.~Brumberg, ``Silent speech interfaces,'' {\em Speech Communication Journal}, vol.~52, no.~4, pp.~270--287, 2010.

\bibitem{maier-hein2005speech}
L.~Maier-Hein, {\em Speech Recognition using Surface Electromyography}.
\newblock Karlsruher Institut f\"{u}r Technologie, 2005.
\newblock Diplom thesis, 121 pages.

\bibitem{ren2023anoverview}
Z.~Ren, K.~Qian, T.~Schultz, and B.~W. Schuller, ``An overview of the icassp special session on ai security and privacy in speech and audio processing,'' in {\em Proc.\ ACM Multimedia workshop}, (Tainan, Taiwan), pp.~1--6, 2023.

\bibitem{diener2015direct}
L.~Diener, M.~Janke, and T.~Schultz, ``Direct conversion from facial myoelectric signals to speech using deep neural networks,'' in {\em Proc.\ IJCNN}, (Killarney, Ireland), pp.~1--7, 2015.

\bibitem{wand2011session}
M.~Wand and T.~Schultz, ``{Session-independent EMG-based Speech Recognition},'' in {\em Proc.\ Biosignals}, (Rome, Italy), pp.~295--300, 2011.

\bibitem{schultz2010modeling}
T.~Schultz and M.~Wand, ``{Modeling coarticulation in EMG-based continuous speech recognition},'' {\em Speech Communication}, vol.~52, no.~4, pp.~341--353, 2010.

\bibitem{scheck2023stream}
K.~Scheck, D.~Ivucic, Z.~Ren, and T.~Schultz, ``{Stream-ETS: Low-latency end-to-end speech synthesis from electromyography signals},'' in {\em Proc.\ Speech Communication, ITG}, (Aachen, Germany), pp.~200--204, 2023.

\bibitem{ren2023self}
Z.~Ren, K.~Scheck, and T.~Schultz, ``Self-learning and active-learning for electromyography-to-speech conversion,'' in {\em Proc.\ Speech Communication, ITG}, (Aachen, Germany), pp.~245--249, 2023.

\bibitem{scheck2023multi}
K.~Scheck and T.~Schultz, ``Multi-speaker speech synthesis from electromyographic signals by soft speech unit prediction,'' in {\em Proc.\ ICASSP}, (Rhodos, Greece), pp.~1--5, 2023.

\bibitem{gaddy2021improved}
D.~Gaddy and D.~Klein, ``An improved model for voicing silent speech,'' in {\em Proc.\ ACL}, (virtual), pp.~175--181, 2021.

\bibitem{jeong2021diff}
M.~Jeong, H.~Kim, S.~J. Cheon, B.~J. Choi, and N.~S. Kim, ``{Diff-TTS: A denoising diffusion model for text-to-speech},'' in {\em Proc.\ INTERSPEECH}, (Brno, Czechia), pp.~3605--3609, 2021.

\bibitem{popov2021grad}
V.~Popov, I.~Vovk, V.~Gogoryan, T.~Sadekova, and M.~Kudinov, ``{Grad-TTS: A diffusion probabilistic model for text-to-speech},'' in {\em Proc. ICML}, (Virtual), pp.~8599--8608, 2021.

\bibitem{choi23d_interspeech}
H.-Y. Choi, S.-H. Lee, and S.-W. Lee, ``{Diff-HierVC: Diffusion-based hierarchical voice conversion with robust pitch generation and masked prior for zero-shot speaker adaptation},'' in {\em Proc.\ INTERSPEECH}, (Dublin, Ireland), pp.~2283--2287, 2023.

\bibitem{ho2020denoising}
J.~Ho, A.~Jain, and P.~Abbeel, ``Denoising diffusion probabilistic models,'' in {\em Proc.\ NeurIPS}, (Vancouver, Canada), pp.~1--12, 2020.

\bibitem{huang2022prodiff}
R.~Huang, Z.~Zhao, H.~Liu, J.~Liu, C.~Cui, and Y.~Ren, ``{Prodiff: Progressive fast diffusion model for high-quality text-to-speech},'' in {\em Proc.\ ACM Multimedia}, (Lisboa, Portugal), pp.~2595--2605, 2022.

\bibitem{kim2022guided}
H.~Kim, S.~Kim, and S.~Yoon, ``{Guided-TTS: A diffusion model for text-to-speech via classifier guidance},'' in {\em Proc.\ ICML}, (Hawaii), pp.~11119--11133, 2022.

\bibitem{kong2020hifi}
J.~Kong, J.~Kim, and J.~Bae, ``{Hifi-GAN: Generative adversarial networks for efficient and high fidelity speech synthesis},'' in {\em Proc.\ NeurIPS}, (Vancouver, Canada), pp.~1--12, 2020.

\bibitem{gaddydigital}
D.~Gaddy and D.~Klein, ``Digital voicing of silent speech,'' in {\em Proc.\ EMNLP}, (Virtual), pp.~5521--5530, 2020.

\bibitem{merlin2020squeeze}
M.~Albes, Z.~Ren, B.~Schuller, and N.~Cummins, ``{Squeeze for sneeze: Compact neural networks for cold and flu recognition},'' in {\em Proc.\ INTERSPEECH}, (Shanghai, China), pp.~4546--4550, 2020.

\bibitem{ren2023fast}
Z.~Ren, T.~T. Nguyen, Y.~Chang, and B.~W. Schuller, ``Fast yet effective speech emotion recognition with self-distillation,'' in {\em Proc.\ ICASSP}, (Rhodes, Greece), 2023.
\newblock 5 pages.

\end{thebibliography}

\end{document}